# SELECTION OF HIGH STRENGTH ENCAPSULANT FOR MEMS DEVICES UNDERGOING HIGH PRESSURE PACKAGING

*Azrul Azlan Hamzah, Yusnira Husaini, Burhanuddin Yeop Majlis, Ibrahim Ahmad*

Institute of Microengineering and Nanoelectronics (IMEN)
Universiti Kebangsaan Malaysia
43600 Bangi, MALAYSIA
e-mail: azlan@vlsi.eng.ukm.my, yusni458@tganu.uitm.edu.my, burhan@vlsi.eng.ukm.my,
ibrahim@vlsi.eng.ukm.my

## ABSTRACT

Deflection behavior of several encapsulant materials under uniform pressure was studied to determine the best encapsulant for MEMS device. Encapsulation is needed to protect movable parts of MEMS devices during high pressure transfer molded packaging process. The selected encapsulant material has to have surface deflection of less than 5 µm under 100 atm vertical loading. Deflection was simulated using CoventorWare ver.2005 software and verified with calculation results obtained using shell bending theory. Screening design was used to construct a systematic approach for selecting the best encapsulant material and thickness under uniform pressure up to 100 atm. Materials considered for this study were polyimide, parylene C and carbon based epoxy resin. It was observed that carbon based epoxy resin has deflection of less than 5 µm for all thickness and pressure variations. Parylene C is acceptable and polyimide is unsuitable as high strength encapsulant. Carbon based epoxy resin is considered the best encapsulation material for MEMS under high pressure packaging process due to its high strength.

## 1. INTRODUCTION

A major ongoing challenge in MEMS is to develop a more standardized packaging like those in IC while maintaining the integrity and functionality of the device. MEMS packaging varies extensively depending on device function, thus driving packaging cost high [1]. In an effort to develop a standardized low cost packaging for MEMS, device capping followed by glob top encapsulation technique is increasingly gaining popularity [2].

Many types of MEMS devices, especially those that do not require interaction with outside ambient, could function well in standardized encapsulation and glob top packaging. In this process, movable parts of a MEMS device are covered by a metal or silicon cap, leaving a gap between the top surface of the movable parts and the bottom surface of the cap. As a result, the cap allows movable elements to move freely while protecting them. The cap is usually fabricated by deposition technique, thus it is not robust enough to endure subsequent high pressure transfer molding process, as the molding pressure could be as high as 100 atm [3]. Therefore, a strong glob top encapsulation is needed to protect the cap, so that the device could be packaged using standardized IC transfer molding process. This paper investigates several materials to select for the best encapsulant under given pressure loading and package thickness constraints.

## 2. ENCAPSULATION FABRICATION PROCESS

A typical MEMS device consists of sensor or actuator elements fabricated on silicon substrate. These elements are usually movable and are very sensitive to damage by chemical contamination, presence of micro dusts, as well as physical touch. MEMS devices such as RF switches, inductors, filters, and accelerometers do not require interaction with outside ambient to function. Hence, a complete isolation of the sensor or actuator elements would increase device performance as well as its lifetime. For this reason, encapsulation is a favorable method for packaging these types of MEMS devices.

Using encapsulation technique, a cap, usually of metal or silicon is deposited on top of MEMS movable elements via deposition processes such as CVD or sputtering [4]. A sacrificial layer, usually of silicon oxide or photoresist material, is pre-deposited on the movable elements to create a gap between the elements and the cap structure. After cap deposition, the sacrificial layer is removed, consequently releasing the movable elements within the enclosed cap. The cap is then sealed by another step of deposition, usually of oxide or metal, thus completely isolating the movable elements from outside ambient.

In order to strengthen the cap structure for the subsequent transfer molding process, an encapsulation layer needs to be added on top of the cap structure as depicted in figure 1. A thin glob is dispensed on capped MEMS device using standard glob top process, yielding encapsulated device with uniform glob top after curing, as depicted in figure 2 below.

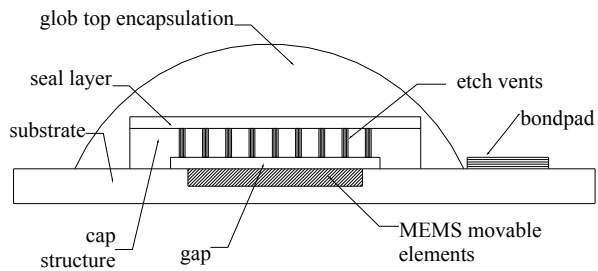

Figure 1. Schematic diagram of a glob top encapsulated MEMS device.

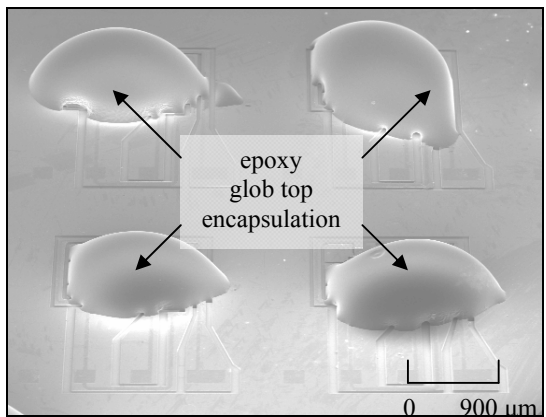

(a)

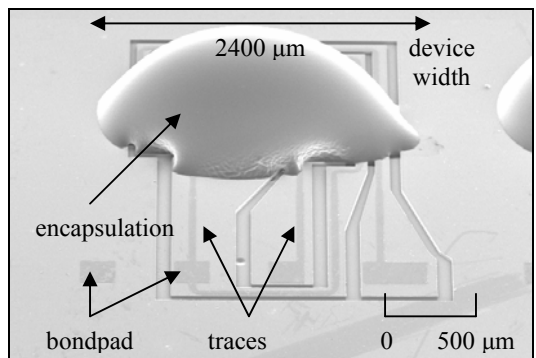

(b)

Figure 2. SEM micrograph showing (a) an array of epoxy resin encapsulated accelerometer devices and (b) a single capped accelerometer device encapsulated with epoxy resin. Note that the traces and bondpads are exposed, while capped accelerometer fingers are enveloped underneath the encapsulation.

## 3. ANALYTICAL APPROACH

A systematic screening approach is applied in encapsulant selection process. The generalized selection process flow is outlined in figure 3 below.

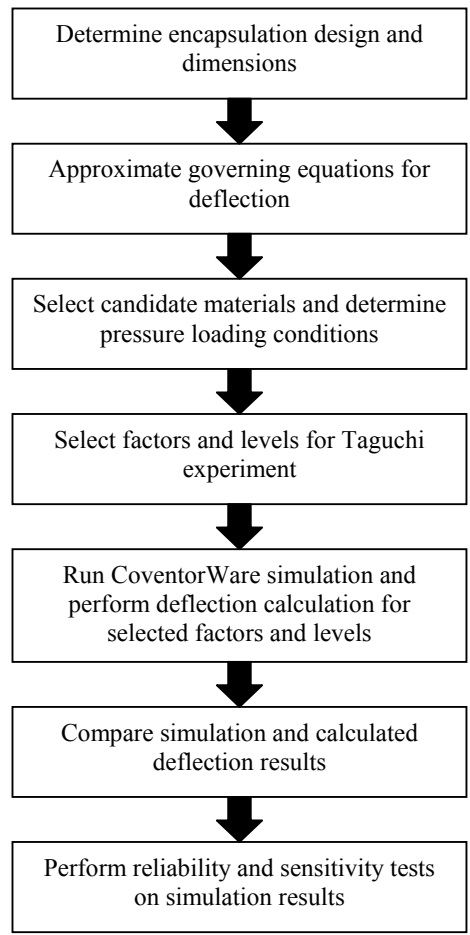

Figure 3. Flow of the encapsulant material selection process.

Initially, dimensional requirements of the encapsulation were determined. For our case, the glob top encapsulation has to have a spherical dome shape with thickness of less than 250 μm (denoted by *t* in figure 4), as the chip would be integrated into thin SMT package. Dome shaped shell was chosen due to its ability to endure high external pressure applied, owing to compressive stress distribution in meridional direction. The lateral length of the glob-top encapsulant, denoted by *2b* in figure 4, should be approximately 2400 μm, limited by device width as shown in figure 2(b). Using the above parameters, the values for shell radius *a* and vertical to base angle *α* could be determined by solving simultaneous equations. The values for *a* and *α* were determined to be 3010 μm and 23.5° respectively.

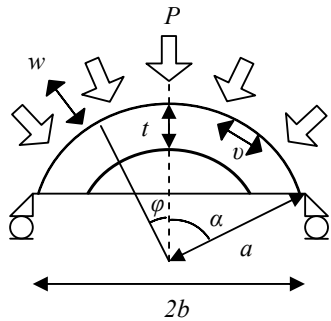

Figure 4. Schematic diagram showing the cross section and the parameters involved in encapsulation structure governing equations.

It is derived from the above parameters that the ratio of shell thickness *t* over curvature *a* is approximately 0.08. Since this ratio is small, it follows that the radial deflection *w* of the encapsulation structure when loaded with uniform force *P* could be approximated using shell bending theory. This theory could be well applied to approximate deflection of thin semi-spherical structure in which the radial deflection is much smaller compared to shell thickness [5]. By means of this theory, radial deflection of a thin shell under uniform loading *P*, as depicted in figure 4, could be formulated as follows [5]:

$$w = \upsilon \cot\varphi - \frac{a^2 P}{Et}\left(\frac{1+\nu}{1+\cos\varphi} - \cos\varphi\right) \quad (1)$$

where *E* and *ν* are the Young's modulus and Poisson ratio of the encapsulant material, *t* is the thickness of the shell, *φ* is the angle from vertical at which the deflection is considered, and *υ* is the meridional deflection, which for a spherical shape can be represented as follows [5]:

$$\upsilon = \frac{a^2 P(1+\nu)}{Et}\left(\frac{1}{1+\cos\alpha} - \frac{1}{1+\cos\varphi} + \ln\left(\frac{1+\cos\varphi}{1+\cos\alpha}\right)\right)\sin\varphi \quad (2)$$

The encapsulant material selected has to be able to withstand up to 100 atm external pressure without excessive deflection. For our capped accelerometer device, the gap between the cap and the device is 5 μm. Therefore, in order to avoid cap material touching the movable parts, maximum allowable deflection on the inner surface of the glob top encapsulant is limited to 5 μm. Materials selected for this study were PMDA polymide, high strength parylene C, and carbon based epoxy resin. These materials were selected for their outstanding mechanical, thermal and chemical resistant properties. Carbon based epoxy resin has the highest Young's modulus, followed by parylene C and polymide. However, parylene C is hermetic, and polymide is more delamination resistant [6]. Young's modulus and poisson ratio values for the selected materials are summarized in table 1 below.

| Material | Young's modulus (GPa) | Poisson ratio |
|---|---|---|
| Polymide | 7.5 | 0.35 |
| Parylene C | 27.59 | 0.4 |
| Carbon epoxy resin | 70 | 0.4 |

Table 1. Young's modulus and poisson ratio values for encapsulant canditate materials [7].

It could be seen from equations (1) and (2) that given the dimensional parameters, Young's modulus, shell thickness, and external pressure applied are the determinant factors for deflection. Poisson ratio on the other hand, is somewhat consistent across the selected candidate materials as shown in table 1. Hence, the material selection criteria have to involve these factors. In order to avoid simulating through the whole spectrum of thickness, pressure, and material selection, Taguchi method was used to construct a systematic approach to screen for the best encapsulant material and thickness under external pressure up to 100 atm. Screening design was used to determine the optimal material and glob top thickness required. The factors selected were Young's modulus, shell thickness, and applied external pressure, and three levels were selected for each factor. Thickness variations considered were 150, 200 and 250 μm and pressure variations were 80, 90 and 100 atm. These values were inserted into L9 orthogonal matrix as shown in table 2, and analyzed by Taguchi approach using JMP software.

| Experiment | Combination | | | Material | Thickness (µm) | P (atm) |
|---|---|---|---|---|---|---|
| 1 | - | + | + | Polyimide | 250 | 100 |
| 2 | + | - | + | C.E.Resin | 150 | 100 |
| 3 | - | 0 | 0 | Polyimide | 200 | 90 |
| 4 | 0 | + | - | Parylene C | 250 | 80 |
| 5 | - | - | - | Polyimide | 150 | 80 |
| 6 | 0 | - | 0 | Parylene C | 150 | 90 |
| 7 | + | 0 | - | C.E.Resin | 200 | 80 |
| 8 | 0 | 0 | + | Parylene C | 200 | 100 |
| 9 | + | + | 0 | C.E.Resin | 250 | 90 |

Table 2. L9 orthogonal matrix showing factor and level combinations for Taguchi experiment.

Simulation study was conducted using CoventorWare ver.2005 software. The glob top encapsulation was approximated as a shelled dome as depicted in figure 4. The simulation model consists of three main parts: silicon substrate, capped MEMS device, and encapsulation shell. A cross-section of modeled encapsulation structure used in simulation is shown in figure 5. Uniform pressure was applied to the dome as to imitate the molding pressure on the encapsulant material. Deflection results obtained from CoventorWare simulation were then compared with deflection results obtained using shell bending equations. Finally, analysis of variance (ANOVA) and sensitivity analysis were conducted on the simulation results to verify the reliability of the statistical model.

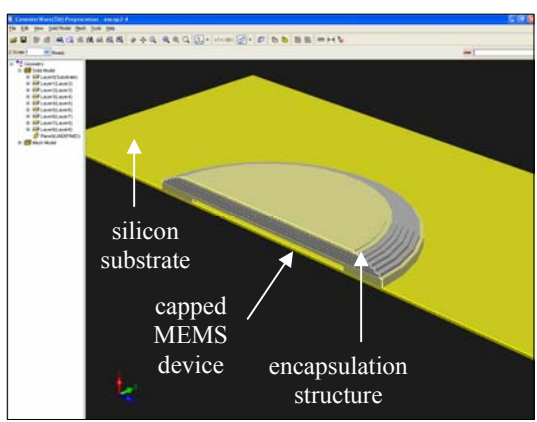

Figure 5. CoventoreWare screen capture of a cross-section view of the encapsulation model used in deflection simulation.

## 4. RESULTS AND DISCUSSION

First and foremost, CoventorWare simulation conducted revealed that maximum deflection occurs at the center of the encapsulation structure. Figure 6 shows a deflected encapsulation structure under uniform external pressure, with major deflection concentrated at the center of the structure. This observation was adapted in deflection calculation using shell bending equations, in which $\varphi$ value is set to approximately zero ($\varphi \approx 0$). This condition was implemented to imitate maximum deflection at the top of the encapsulation structure observed in CoventorWare simulation.

Surface deflection values obtained from simulation performed with the aforementioned materials, pressure, and thickness variations are tabulated in table 3 below. Alongside the simulation results, deflection values calculated using shell bending equations (1) and (2) above are also tabulated for comparison. It could be seen that the calculated and simulated results are close most of the times, but deviate profusely at some instances. However, one could observe that the deviation between the two sets of results is generally less than 38%. Therefore, the calculated results provide a good basis for simulation verification. Based on both sets of results, it could be concluded that the simulated deflection values are in close proximity to the actual deflection values given the parameters and loading conditions stated above.

| Exp | Material | Thickness (µm) | P (atm) | $w_s$ sim (µm) | $w_c$ calc (µm) | Error $w_c/w_s$ (%) |
|---|---|---|---|---|---|---|
| 1 | Polyimide | 250 | 100 | 12.59 | 15.58 | 23.79 |
| 2 | C.E.Resin | 150 | 100 | 4.95 | 3.11 | -37.25 |
| 3 | Polyimide | 200 | 90 | 18.94 | 17.53 | -7.42 |
| 4 | Parylene C | 250 | 80 | 2.60 | 3.12 | 20.07 |
| 5 | Polyimide | 150 | 80 | 33.18 | 20.78 | -37.38 |
| 6 | Parylene C | 150 | 90 | 2.93 | 5.85 | 99.78 |
| 7 | C.E.Resin | 200 | 80 | 1.70 | 1.54 | 9.44 |
| 8 | Parylene C | 200 | 100 | 5.39 | 4.88 | 9.44 |
| 9 | C.E.Resin | 250 | 90 | 1.16 | 1.38 | 19.69 |

Table 3. Comparison of simulated and calculated deflection values for all combinations tested.

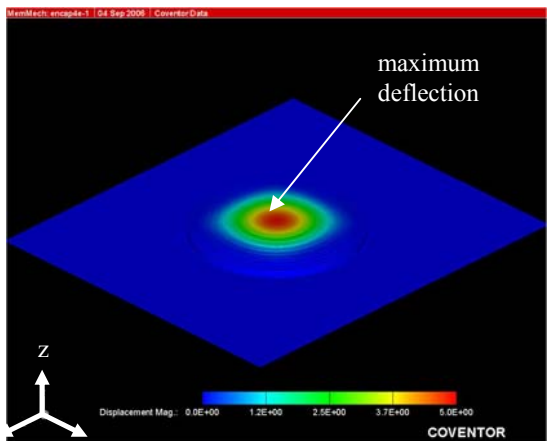

Figure 6. A simulation capture showing 150 μm thick carbon based epoxy resin with deflection of 4.95 μm under 100 atm pressure.

The deflection values obtained from CoventorWare simulation were analyzed using Taguchi method. Figure 7 shows JMP results that relate encapsulation deflection values with each factor considered. Factor desirability is plotted underneath deflection values of each factor at each level, where desirability increases with decrease in deflection value. The vertical lines mark level selected for each factor. The top horizontal lines mark the corresponding deflection for the selected factor combination. On the other hand, the bottom horizontal lines mark the desirability of each factor at the level selected. The intersection of the horizontal line and the slanted curve in the utmost right column marks the total desirability of the level combination. The slanted dashed curves at the very top of both second and third columns indicate the range of continuous thickness and pressure variation. The JMP curves are used to determine optimal encapsulation thickness given the deflection limit. Optimized result in figure 7(a) shows that 172.5 μm thick carbon based epoxy resin glob top would deflect 4.09 μm under 100 atm pressure. Similarly, parylene C glob top of thickness 205 μm would deflect 4.98 μm under 100 atm pressure (figure 7(b)). Deflection curves for polymide are not shown as it deflects more than 5 μm for any encapsulation thickness and pressure combination.

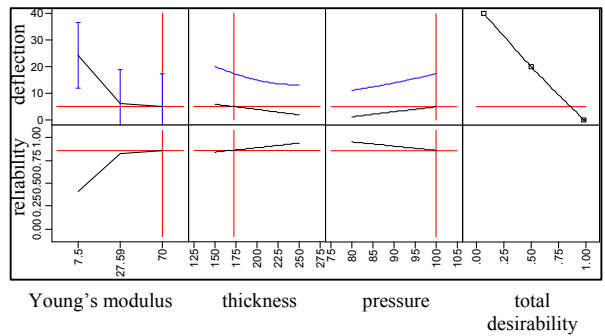

(a)

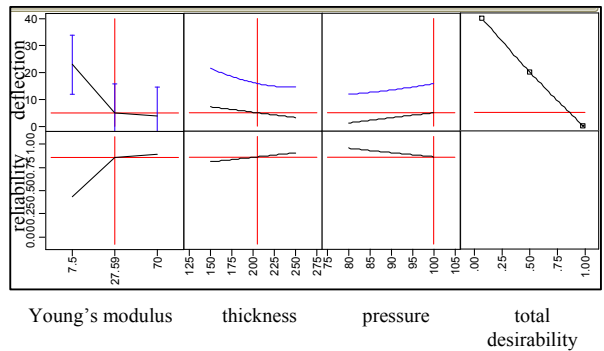

(b)

Figure 7. JMP results showing deflection values for (a) 172.5 μm thick carbon based epoxy resin under 100atm pressure. (b) 205 μm thick parylene C under 100atm pressure. Note that the deflection values are less than 5 μm for both cases.

ANOVA was used to evaluate the adequacy of the Taguchi model developed. Table 4 summarizes ANOVA results. The P-value indicates that the model is more than 95% significant in the study of surface deflection, which statistically proves the high precision of the results.

| Source | df | SS | MS | F | P-value |
|---|---|---|---|---|---|
| Simulation | 4 | 1443.2734 | 360.818 | 7.5094 | 0.0242 |
| Errors | 5 | 240.2443 | 48.049 | | |
| Total | 9 | 1683.5177 | | | |

Table 4. ANOVA test results on the adequacy of the Taguchi model developed. The parameter df is degree of freedom, SS is sum of square, MS is mean of square, F is SS divided by MS, and P-value is the smallest alpha.

Another important aspect to consider is the sensitivity of each factor in the model. The main effect test was used to check the sensitivity of each factor towards surface deflection. In this test, P-value indicates the influence of a particular factor on surface deflection. A higher P-value means that a factor is less sensitive. Table 5 summarizes the main effect test results. It is observed that Young's modulus is the most sensitive factor in the model, compared to encapsulation thickness and applied pressure, indicated by its small P-value. Thus, a small change in Young's modulus value would greatly affect surface deflection. It is therefore most crucial to choose a material with correct Young's modulus value in order to obtain the desired deflection result.

| Source | Nparm | df | SST | F | P-value |
|---|---|---|---|---|---|
| Young's Modulus | 2 | 2 | 694.1383 | 7.2233 | 0.0335 |
| Thickness | 1 | 1 | 28.7176 | 0.5977 | 0.4744 |
| Pressure | 1 | 1 | 123.2013 | 2.5641 | 0.1702 |

Table 5. Main effect test results. The parameter Nparm is number of parameters, df is degree of freedom, SST is total sum of square, F is F distribution, and P-value is the smallest alpha.

From the simulation, calculation, and analyses conducted, the best material for MEMS encapsulation and the important factors involved have been successfully analyzed. Carbon based epoxy resin was determined the best material for encapsulation of MEMS devices undergoing high pressure packaging.

## 5. CONCLUSIONS

CoventorWare is an excellent tool for simulating deflection of encapsulation surface under external pressure. The simulated deflection values are in close proximity to deflection values obtained using shell bending equations. Screening design has effectively simplified the encapsulant selection process. Carbon based epoxy resin and parylene C are acceptable glob top materials since their deflection under 100 atm loading are less than 5 μm for thickness within 250 μm limit. Polyimide is deemed unsuitable since its deflections are greater than 5 μm for the entire thickness and pressure variations. Carbon based epoxy resin was selected as the best encapsulation material for MEMS devices undergoing high pressure packaging process due to its high strength.